\documentclass[aps,showpacs,amsmath,amssymb]{revtex4}
\usepackage{graphicx}

\usepackage{graphicx}% Include figure files

\usepackage{dcolumn}% Align table columns on decimal point
\usepackage{epsfig}% bold math
\usepackage{verbatim}

\def\be{\begin{equation}}
\def\ee{\end{equation}}
\newcommand{\ba}{\begin{array}}
\newcommand{\ea}{\end{array}}
\newcommand{\bea}{\begin{eqnarray}}
\newcommand{\eea}{\end{eqnarray}}

\newcommand{\op}[1]{\ensuremath{\widehat{ #1} }}
\newcommand{\der}{\partial}
\newcommand{\vct}[1]{\ensuremath\mbox{\boldmath$ #1 $}}
\newcommand{\mat}[1]{\ensuremath\mbox{$ \mathbb #1 $}}

\newcommand{\HH}{ \mathcal H}
\newcommand{\GO}{ \mathcal O}
\newcommand{\N}{ \mathcal N}
\newcommand{\vx}{\vct x}
\newcommand{\vy}{\vct y}

\newcommand{\vxi}{\vct \xi}

\newcommand{\vl}{\vct l}

\begin{document}

\title{Complex WKB Evolution of Markovian Open Systems} 

\author{O.~Brodier$^{1}$ and A.~M.~Ozorio de Almeida$^{2}$}
\affiliation{$^1$ Laboratoire de Math\'ematiques et Physique Th\'eorique,
Facult\'e des Sciences et Techniques, Universit\'e de Tours, 37200 TOURS \\
$^2$ Centro Brasileiro de Pesquisas F\'isicas, 
Rua Dr. Xavier Sigaud, 150, 22290-180 Rio de Janeiro, BRASIL}

\date{\today}

\begin{abstract}

We derive a semiclassical approximation for the evolution generated by the Lindblad equation 
as a generalization of complex WKB theory. Linear coupling to the environment
is assumed, but the Hamiltonian can be a general function of positions and momenta.  
The theory is carried out in the chord representation 
and describes the evolved quantum characteristic function, 
which gives direct access to the Wigner function 
and the position representation of the density operator by Fourier transforms. 
The propagation is shown to be of Liouville type in a complex double phase space, 
the imaginary part of the action being responsible for decoherence. 
The theory is exact in the quadratic case, just as the real WKB theory 
that we previously developed for the Markovian evolution of extended states, 
but it also describes the decoherent and dissipative evolution of localized states, 
such as the interference terms of a Schr\"odinger cat state. 
The present rederivation of the real WKB approximation leads to its interpretation 
as a first order classical perturbation of the complex theory and to a discussion of its validity.
The example of a simple cubic Hamiltonian illustrates the various levels of approximation 
derived from the complex WKB theory.

\end{abstract}


\maketitle

\section{Introduction}

The general state of a system in quantum mechanics can be fully described by the density operator $\op \rho$, also called the ``state operator'', normalized by $\mathop{Tr}\op\rho = 1$. If the system is known to be in a ``pure state'', that is, a single vector of the corresponding Hilbert space, then $\op\rho$ is the projector on this vector, $\op \rho = |\psi\rangle\langle\psi|$, and then its {\it Von Neumann entropy}, $\mathop{S} = -\mathop{Tr}\op\rho\log\op\rho$, is equal to $0$, and its {\it purity}, $\mathop{Tr}\op\rho^2$, is equal to $1$. However, the density operator generalizes this case in that it can also describe a statistical mixture of such states, which expresses a partial knowledge of the system. Then it is not a projector any more but a sum of several projectors, and one has $S>0$ and $\mathop{Tr}\op\rho^2 < 1$.

When the system is isolated, the evolution is unitary and preserves the purity of the state. However, when the system interacts with its environment, which is always the case in realistic situations, an initially pure state will undergo a non-unitary evolution which will not preserve purity: one is in the paradigm of quantum open systems. The consequent loss is two-fold. {\it Decoherence} corresponds to the vanishing of the off-diagonal terms of the density operator, leading finally to a statistical distribution which can be interpreted classically. Thus it can dynamically explain one aspect of the postulate of reduction of the wave packet. {\it Dissipation} corresponds to a loss or gain of energy, and is usually considered as much slower than decoherence, especially for large systems.

The general evolution of $\op\rho(t)$ for an open system is a practical issue in many experimental situations, already well established in quantum optics\cite{Leon97}, also in atomic and nuclear physics, and more recently in the physics of quantum information where it is of crucial importance to controle the interaction of the system with its environment, in order to avoid decoherence. To construct theoretical models with the maximum of generality, it is generally admitted that such an evolution should satisfy at least three basic requirements: preservation of Hermiticity ($\mathop{Tr}\op\rho\op A\in{\mathcal R}$), trace ($\mathop{Tr}\op\rho=1$) and positivity ($\mathop{Tr}\op\rho|\psi\rangle\langle\psi|\geq 0$). 

Two main approaches have been explored to obtain explicit dynamics of an open quantum system. The first is to consider that the open system is part of a bigger closed system obeying a unitary evolution, and to trace over the environment part, that is over all the degrees of freedom which are not directly concerned with the system under study. Such an approach leads to the Bloch-Redfield equations \cite{Blo57}\cite{Red57}. The second is to assume that, beyond the above general requirements, the time evolution should obey a semigroup law, that is, a forward translational time invariance. Then the most general equation was derived by Lindblad\cite{Lin76} (see also \cite{Isa94}). 

The Lindblad equation can be written 
\begin{equation}
\frac{\der \op\rho}{\der t} = -\frac{i}{\hbar}\Big[ \op H, \op\rho \Big]
+\frac{1}{2 \hbar}\sum_k\left(
2 \op L_k\op\rho\op L_k^\dagger - L_k^\dagger\op L_k\op\rho - 
\op\rho L_k^\dagger \op L_k
\right).
\end{equation}
The Hamiltonian $\op H$ describes the unitary evolution of the system without environment. There is no restriction on the operators $\op L_k$, commonly dubbed Lindblad operators, and they are not Hermitian when the coupling to the environment is dissipative. For instance, the master equation of quantum optics can be seen as a special case of the Lindblad equation, with the operator $\op L_1 = \op a$, i.e. the annihilation operator, describing the emission process, and $\op L_2 = \op a^\dagger$ the absorption. 

In this paper, we will focus on finding semiclassical approximate solutions of a Lindblad equation with a general Hamiltonian part and Lindblad operators that are linear functions of momentum and position. 
This involves a considerable adaptation of the more familiar semiclassical approximations for pure states. 
In previous papers, we generalized the analysis of the evolution of extended states, 
the WKB theory based on a real classical phase space, that goes back to Van Vleck \cite{Van Vleck} 
(see also \cite{Gutzwiller}\cite{Maslov}\cite{Almeida}). 
Here we develop the WKB theory on a complexified phase space, thus generalizing previous semiclassical
theory of unitary evolution developed by Huber, Heller and Littlejohn \cite{Hub03} and by Maslov \cite{maslov:book2}.

The advantage of the complex theory is that 
it allows us to include in the same theory density operators that are localized, such as, typically, the interference terms arising from superposition of coherent states. Also it is more robust to the fast damping of quantum correlations induced by decoherence, which breaks down the validity of the real stationary phase method.

Just as with pure states, the evolution may in time extend the state beyond the range of validity for the theory
and this intermediate stage also demands attention (see e.g. \cite{Vall08}, for pure state evolution).
However, this is a case where the treatment of open systems has an advantage: usually the process of decoherence is exponentially fast with respect to other relevant features of the motion.   

The complex WKB theory can also be used to rederive its real counterpart.
Both become exact in the special case of a quadratic Hamiltonian and Lindblad operators that are linear
in the positions and momenta. However, we find that the complex theory is  more accurate,
even for extended states for which the real WKB theory should hold. Fortunately, we are able to show
that the perturbation theory that bridges both theories is farely accurate, if one uses the proper initial condition, which is fulfilled by a generic propagator introduced in \cite{AlmBro07}.
Therefore, it is not necessary to abandon the more transparent intuitive content of real WKB 
for the evolution of extended states.

We represent the solution in the ``chord'' space, 
that is the Fourier conjugate of the Weyl-Wigner representation of the state operator, 
which is defined in the ``centre'' space. Together they describe a {\it double phase space}, 
where the {\it double Hamiltonian} generates corresponding classical trajectories 
that support the semiclassical approximation. 
It is this double phase space that is now complexified.
The present treatement is in continuity of a series of papers, 
\cite{BroAlm04}\cite{AlmBro06} and \cite{AlmBro07}, but it can be understood separately, 
since all the approximations are quite transparent. 

We first give a reminder for the Weyl-Wigner representation, and write the Lindblad equation in this formalism. Then we describe the general WKB procedure by introducing an adapted complex double phase space driven by a complex Hamiltonian, which leads to a Hamilton-Jacobi equation. After verifying agreement of the theory with the quadratic case, we develop an effective real WKB dynamics, based on a perturbative approach, which is valid for states whose initial classical action is real. We apply this real WKB to a generic mixed propagator $R_{\vct x}(\vct y,t)$ which can turn an initial Wigner function $W_0(\vct x)$ into the evolving chord function $\chi(\vct y,t)$. Finally we construct a simple non-quadratic example where all the methods are compared.

%{\it descricao das secoes?}

One should note that all the following formulae are appropriate for a system with a single degree of freedom, in order to clarify the notation. Nonetheless, it is quite simple to generalize our treatment and results for a finite number of degrees of freedom.

\section{Lindblad equation in the Weyl formalism}

The Weyl representation maps every quantum operator onto a phase space function, that is, a function of the vector $\vx=(p,q)$\cite{Wey31}\cite{Alm98}. For an operator $\op A$ the Weyl symbol $A$ is defined as
\begin{equation}
A(\vx)= 2 \int \exp{(-\frac{i}{\hbar}p Q )}
~ \langle q +\frac{Q}{2}| \op A | q - \frac{Q}{2}\rangle~d Q.
\end{equation}
The Weyl symbol of the state operator $\op\rho$ is the Wigner function
\begin{equation}
W(\vx) = \N\int \exp{(-\frac{i}{\hbar}pQ )}
~\langle q+\frac{Q}{2}|\op \rho|q-\frac{Q}{2}\rangle ~dQ,
\end{equation}
with $\N = 1/(2\pi\hbar)$; and its Fourier transform, the chord function $\chi(\vct \xi)$, also called characteristic function, is
\begin{equation}
\chi(\vct \xi) = \N \int 
\exp{(-\frac{i}{\hbar}\vct \xi\wedge \vct x )}
~W(\vct x)~d\vct x,
\label{fourierwigner}
\end{equation}
where the wedge product of two vectors $\vct x=(p,q)$ and $\vct x'=(p',q')$ is defined by $\vct x\wedge \vct x'=pq'-p'q = \mat J \vct x \cdot \vct x'$, which also defines the skew matrix $\mat J = \left(\begin{array}{cc} 0 & -1 \cr 1 & 0 \end{array}\right)$. One can have the chord symbol directly from the quantum operator through the formula
\begin{equation}
\chi(\vct \xi) = \N \int 
\exp{(-\frac{i}{\hbar}\xi_p q)}
~\langle q+\frac{\xi_q}{2}|\op \rho|q-\frac{\xi_q}{2}\rangle ~dq.
\label{chordfunc}
\end{equation}
We call the space of all $\vx$ the centre space, and the space of all
$\vct\xi$ the chord space. 

In the chord space, by using product rules for the product of operators\cite{Alm98}, the Lindblad equation is represented by a partial differential equation. This equation is actually simpler than in the Weyl (centre) representation, and this justifies our choice. In the case where the Lindblad operators are linear functions of $\op p$ and $\op q$, that is, $\op L = \vct l'\cdot\op{\vct x} + i\vct l''\cdot\op{\vct x}$ with $\vct l'$ and $\vct l''$ real vectors, this equation can be written
\begin{eqnarray}
\frac{\der \chi}{\der t} (\vct \xi,t) & = & -\frac{i}{\hbar}
\N \int \Bigl[H(\vx'+\frac{1}{2}\vct \xi,t)
-H(\vx'-\frac{1}{2}\vct \xi,t)
\Bigr]
~\exp{ \left(\frac{i}{\hbar}(\vxi'-\vxi)\wedge \vx' \right)}
~\chi(\vct \xi',t)
~d\vct \xi'~d\vx' \cr
~ & ~ & -\gamma ~\vct \xi\cdot \frac{\der \chi}{\der \vct\xi}(\vct \xi,t) 
- \frac{1}{2\hbar}~\Bigl[(\vl'\cdot\vct \xi)^2 + (\vl''\cdot\vct \xi)^2 \Bigr] 
~\chi(\vct \xi,t). 
\label{dynchi}
\end{eqnarray}
The {\it dissipation coefficient}, 
\begin{equation}
\gamma = \vl''\wedge\vl',
\end{equation}
is null for a Hermitian Lindblad operator ($\vl''=\vct 0$) and we then have a purely diffusive case. $H$ is the Weyl representation of the Hamiltonian of the isolated system and coincides with the corresponding classical Hamiltonian, up to corrections coming from non-commutativity of $\op p$ and $\op q$. Its arguments in equation (\ref{dynchi}) are the pair of remarkable points $\vct x_+ = \vct x + \frac{\vct \xi}{2}$ and $\vct x_- = \vct x - \frac{\vct \xi}{2}$, which can be considered as both tips of a chord $\vct \xi$. Although this chord $\vct \xi = (\vct\xi_p,\vct\xi_q)$ can be interpreted as an auxiliary conjugate variable of $\vct x$, in the current approach it is actually more convenient to write the solution in terms of $\vct y = \mat J\vct\xi = (-\vct\xi_q,\vct\xi_p)$. Indeed, the direct sum of these conjugate spaces can be interpreted as a double phase space, where $\vct x$ formally plays the role of the position $q$, and $\vy$ the role of its Fourier conjugate, the momentum $p$. Then the above equation becomes
\begin{eqnarray}
\frac{\der \chi}{\der t} (\vct y,t) & = & -\frac{i}{\hbar} 
\N \int \Bigl[H(\vx'- \frac{1}{2}\mat J\vct y ,t)
-H(\vx' + \frac{1}{2}\mat J\vct y,t)
\Bigr] 
~\exp{ \left(\frac{i}{\hbar}(\vct y'- \vct y)\cdot \vx' \right)}
~\chi(\vct y',t)
~d\vct y'~d\vct x' \cr
~ & ~ &
-\gamma ~\vct y \cdot \frac{\der \chi}{\der \vct y}(\vct y,t) 
- \frac{1}{2\hbar}\;~\Bigl[(\vct\lambda'\cdot\vct y)^2 + (\vct\lambda''\cdot\vct y)^2 \Bigr]\chi(\vct y,t). 
\label{dynchi_y}
\end{eqnarray}
The same name has been kept for the characteristic function $\chi(\vct y,t)$, 
though strictly this should be $\chi(\vct \xi,t)=\chi(-\mat J \vct y,t)$, 
and we have set the complex vector
\begin{equation}
\vct\lambda= \vct \lambda' + i\vct \lambda'' ={\mat J} (\vct l' + i\vct l'').
\label{lamb}
\end{equation}
The term $\vct y \cdot \frac{\der \chi}{\der \vct y}$ can actually be included in the integral term, thanks to an integration by parts of the exponential, and one has finally
\begin{eqnarray}
\frac{\der \chi}{\der t} (\vct y,t) = -\frac{i}{\hbar} 
\N \int \HH (\vct x',\vct y,t)
~\exp{ \left(\frac{i}{\hbar}(\vct y'- \vct y)\cdot \vx' \right)} 
~\chi_t(\vct y')~d\vct y'~d\vct x'\cr 
- \frac{1}{2\hbar}\;~\Bigl[(\vct\lambda'\cdot\vct y)^2 + (\vct\lambda''\cdot\vct y)^2 \Bigr]
~\chi(\vct y,t), 
\label{dynchi_synth}
\end{eqnarray}
with
\begin{eqnarray}
\HH(\vct x,\vct y,t) & = & H(\vct x - \frac{1}{2}\mat J \vct y,t)
-H(\vct x + \frac{1}{2}\mat J \vct y,t) 
- \gamma ~ \vct x\cdot\vct y \cr
~ & = & \HH^+(\vct x,\vct y,t) - \HH^-(\vct x,\vct y,t) - \gamma ~ \vct x\cdot\vct y.
\label{defHH}
\end{eqnarray}
This is exactly the {\it double Hamiltonian} that generates the classical motion underlying the semiclassical approximations in \cite{AlmBro07}. Obviously, the double Hamiltonian will be time-independent if it is obtained from a time-independent single Hamiltonian. Furthermore, in the absence of dissipation, both 
$\HH^+(\vct x,\vct y)=H(\vct x - \frac{1}{2}\mat J \vct y)$ and
$\HH^-(\vct x,\vct y)=H(\vct x + \frac{1}{2}\mat J \vct y)$ 
will also be constants which generate independent motions for both chord tips.

Notice that (\ref{dynchi_synth}) can also be written as
\begin{eqnarray}
\frac{\der \chi}{\der t} (\vct y,t) = -\frac{i}{\hbar} 
\N \int \HH (\vct x',\vct y,t) 
~\exp{ (-\frac{i}{\hbar}\vct y\cdot \vx' )}
~W_t(\vct x')~d\vct x'\cr 
- \frac{1}{2\hbar}\;~\Bigl[(\vct\lambda'\cdot\vct y)^2 + (\vct\lambda''\cdot\vct y)^2 \Bigr] 
~\chi(\vct y,t), 
\label{dynchi_synth_W}
\end{eqnarray}
in terms of the evolving Wigner function. Alternatively, the definition of 
the {\it complex double Hamiltonian}
\begin{equation}
\HH_c (\vct x,\vct y,t)= \HH (\vct x,\vct y,t)- \frac{i}{2}\;
~\Bigl[(\vct\lambda'\cdot\vct y)^2 + (\vct\lambda''\cdot\vct y)^2 \Bigr],
\label{compdouble}
\end{equation}
leads to
\begin{equation}
\frac{\der \chi}{\der t} (\vct y,t) = 
-\frac{i}{\hbar}
\HH_c \left(-\frac{\hbar}{i}\frac{\der}{\der \vct y}^{(1)},\vct y^{(2)},t\right)
~\chi(\vct y,t), 
\label{dynchi_der_H}
\end{equation}
where $^{(1)}$ and $^{(2)}$ mean that the derivatives are taken first and then the $\vct y$ terms are multiplied.
Given this specific choice of ordering for the operators 
$\widehat{\vct y}=\vct y$ and $\widehat{\vct x}=-\frac{\hbar}{i}\frac{\der}{\der \vct y}$, 
this is completely analogous to the Schrodinger equation, 
which thus allows us to extend the various forms of WKB theory, once it is recalled that 
variations in operator ordering have effects that are semiclassically small.

The differential term in the RHS of (\ref{dynchi_der_H}) 
(or the integral term in the RHS of (\ref{dynchi_synth_W}))  
represents the unitary part of the evolution. In other words,
in the chord representation, the commutator is specified by the real double Hamiltonian as
\begin{equation}
\Bigl[\op H,\op \rho\Bigr]_{\textrm chord}
%\HH \left(\frac{\hbar}{i}\frac{\der}{\der y_p}^{(1)},
%\frac{\hbar}{i}\frac{\der}{\der y_q}^{(1)},
% y_p^{(2)}, y_q^{(2)}, t\right) 
%\chi(\vct y,t)
=\HH \left(-\frac{\hbar}{i}\frac{\der}{\der \vct y}^{(1)},
\vct y^{(2)}, t\right) \;\chi(\vct y,t).
\label{unitary}
\end{equation}

\section{General Complex Dynamics}
\label{complexd}

We here assume that the chord representation $\chi(\vct y,t)$ of the localized state has the usual semiclassical form 
\begin{equation}
\chi(\vct y,t) = \exp{\frac{i}{\hbar}S(\vct y,t)},
\label{simple}
\end{equation}
where $S(\vct y,t)$ is a function with complex values of order $\GO(\hbar^0)$. One shall be aware that, if we find the time evolution of such a state determined by the Lindblad equation, then we can also evolve any linear combination of such states, which can be coherent states for instance. This is a consequence of the linearity of the Lindblad equation.

This semiclassical form naturally induces a $\hbar$ expansion of the unitary part of the equation, as it is shown in the appendix \ref{app_devH},
\begin{equation}
\HH \left(-\frac{\hbar}{i}\frac{\der}{\der \vct y}^{(1)},
 \vct y^{(2)}, t\right) ~\chi(\vct y,t) = 
\Biggl[
\HH \left(-\frac{\der S}{\der \vct y}(\vct y,t), \vct y, t\right) + \GO(\hbar)
\Biggr] ~\chi(\vct y,t).
%\frac{\hbar}{i}\frac{\der \HH}{\der \vct x}\cdot \frac{\der}{\der \vct y} \ln{\phi(\vct y)} + \frac{\hbar}{2i}\mathop{Tr}\left(\frac{\der^2 \HH}{\der \vct x^2}\frac{\der^2 S}{\der \vct y^2}\right)
\end{equation}
Hence, by expanding (\ref{dynchi_der_H}) at leading order in $\hbar$, one obtains
the Hamilton-Jacobi equation
\begin{eqnarray}
\frac{\der S}{\der t}(\vct y,t)  = 
-\HH\left( -\frac{\der S}{\der \vct y}(\vct y,t), \vct y ,t\right)  
+ \frac{i}{2} ~\Bigl[(\vct\lambda'\cdot\vct y)^2 + (\vct\lambda''\cdot\vct y)^2 \Bigr] + \GO(\hbar) \cr
=-\HH_c\left( -\frac{\der S}{\der \vct y}(\vct y,t), \vct y ,t\right)
+ \GO(\hbar).
\label{eq_ordre1_complex}
\end{eqnarray}

This leads to a double phase space generalization of the complex WKB theory in \cite{Hub03}.
Notice that the present Hamilton-Jacobi equation is defined by a Hamiltonian
of unusual form: not only does it not separate into the familiar kinetic and potential energy terms,
but it is complex. Therefore the action, $S(\vct y,t)$, becomes complex, 
even in the special case where the initial action,
$S_0(\vct y)=S(\vct y,0)$ is real. Because of these unusual features, we will work directly
with the Hamiltonian formalism in the full complex double phase space, 
rather than attempting to connect the evolving action to a Lagrangian. 

In general the initial action specifies a {\it Lagrangian surface}
of half the dimension of the complex double phase space,
\begin{eqnarray}
\vct y_0 & = & \vct z \cr
\vct x_0 & = & -\frac{\der S_0}{\der \vct y}(\vct z).
\label{Lagman}
\end{eqnarray}
To obtain the formal solution of the Hamilton-Jacobi equation, it is convenient to define a family of classical trajectories $(\vct x_t,\vct y_t)$ in the complex double phase space with initial conditions on this initial surface
and driven by the complex Hamiltonian (\ref{compdouble}) through Hamilton's equations:
\begin{eqnarray}
\dot{\vct y}_\tau & = & -\frac{\der \HH_c}{\der \vx}(\vct x_\tau,\vct y_\tau,\tau)
=-\frac{\der \HH}{\der \vx}(\vct x_\tau,\vct y_\tau,\tau)\cr
\dot{\vct x}_\tau & = & \frac{\der \HH_c}{\der \vy}(\vct x_\tau,\vct y_\tau,t)
=\frac{\der \HH}{\der \vy}(\vct x_\tau,\vct y_\tau,t) 
- i(\vct\lambda'\cdot\vct y_\tau)\; \vct\lambda'-i(\vct\lambda''\cdot\vct y_\tau)\; \vct\lambda''.
\label{Hamilton} 
\end{eqnarray}
%{\it superficie inicial e' 2D, com t temos 3D}
In the present case of a single degree of freedom, this family is spanned by variables $\vct z$ ($2$ complex dimensions) and $\tau$ ($1$ real dimension). Hence, considering the real and imaginary parts of each complex variable,
this forms a real submanifold $\mathcal M$ with $5$ dimensions, within a real phase space of $8$ dimensions,
corresponding to the $4$ complex dimensions of $(\vct x,\vct y)$. This submanifold, which will serve as a backbone to build our solution, is completely parametrized by $(\vct y,\tau)$ through a function $\vct x_{\mathcal M}(\vct y,\tau)$, such that $\left(\vct x_{\mathcal M}(\vct y,\tau),\vct y\right)$ is the most general point of $\mathcal M$.

In the real case, it is a well known fact, see for instance \cite{Arnold:book}, that the solution $S(\vct y,t)$ of the Hamilton-Jacobi equation
\begin{equation}
\frac{\der S}{\der t}(\vy,t) = - \HH\left( -\frac{\der S}{\der \vy}(\vct y,t),\vct y ,t\right)
\label{Ssolut}
\end{equation}
is the generating function of the submanifold $\left(\vct x_{\mathcal M}(\vct y,t),\vct y\right)$ for each $t$. 
The complex generalization is straightforward and is proved in Appendix \ref{app_closedform}
by showing that the differential form
\begin{equation}
\delta s = -\vct x_{\mathcal M}(\vct y,\tau) \cdot d\vct y - \HH_c(\vct x_{\mathcal M}(\vct y,\tau),\vct y,\tau)d\tau 
\end{equation}
is a closed form on $(\vy,\tau)$. Hence we can define the solution
\begin{equation}
S(\vy,t) = S_0(\vy_i) + \int_{\vy_i,0}^{\vy,t} \delta s,
%\vct x_{\mathcal M}(\vct y',\tau) 
%\cdot d \vct y'
%- \HH\left( \vct x_{\mathcal M}(\vct y',\tau), \vct y' ,\tau\right) ~d\tau,
\label{S1}
\end{equation}
where $\vy_i$ is arbitrary and the integral can be performed along any path of $\mathcal M$ 
joining $(\vy_i,0)$ to $(\vy,t)$. 

A natural choice of such a path is $(\bar{\vct y}_\tau,\tau)$, such that $(\bar{\vct x}_\tau,\bar{\vct y}_\tau)$ is the classical trajectory of $\mathcal M$ with $\bar{\vct y}_t=\vct y$, litterally the history of $\vct y$ at time $t$. We have in particular $\vct x_{\mathcal M}(\bar{\vct y}_\tau,\tau) = \bar{\vct x}_\tau$ by construction. The choice of this trajectory sets the value of $\vct y_i = \bar{\vct y}_0$. Then one can write more explicitely
\begin{equation}
S(\vy,t) = S_0(\bar{\vct y}_0) + \int_0^t 
\Bigl[ 
- \bar{\vct x}_\tau \cdot \frac{\der \bar{\vct y}_\tau}{\der \tau}
- \HH_c\left( \bar{\vct x}_\tau,\bar{\vct y}_\tau,\tau\right)\Bigr]~d\tau.
\label{S}
\end{equation}
One should keep in mind that $(\bar{\vct x}_\tau,\bar{\vct y}_\tau)$ implicitly depends on $(\vct y,t)$, and in particular $\bar{\vct y}_0$ and $\bar{\vct x}_0=-\frac{\der S_0}{\der \vct y}(\bar{\vct y}_0)$ are complex functions of $(\vct y,t)$. There may be many complex trajectories from the initial surface to that with fixed $\vct y$, just as in ordinary complex WKB for unitary evolution \cite{Hub03}. However, in the case that the initial action is real 
and the Lindblad terms are small, the relevant trajectory should lie close to a real trajectory of the unitary problem. This motivates a perturbative real WKB theory introduced in a further section.
Tunneling might be included in the underlying classical description by allowing complex time trajectories.

The final expression of the solution, in terms of the real part of the double Hamiltonian, is then
\begin{equation}
\chi (\vy,t) = {\mathcal K}~\exp{\left( 
 \frac{i}{\hbar} S_0(\bar{\vct y}_0) -
\frac{i}{\hbar} \int_0^t \Bigl[\bar{\vct x}_\tau \cdot\frac{\der \bar{\vct y}_\tau}{\der \tau}
+ \HH\left( \bar{\vct x}_\tau,\bar{\vct y}_\tau,\tau\right)\Bigr]~d\tau
- \frac{1}{2\hbar}\int_0^t ~\Bigl[(\vct\lambda'\cdot\bar{\vct y}_\tau)^2 + (\vct\lambda''\cdot\bar{\vct y}_\tau)^2 \Bigr]
 ~d\tau \right)}.
\label{solutchigen}
\end{equation}

\section{The quadratic case}

Here we show that in the quadratic case, the solution (\ref{solutchigen}) of the complex WKB approximation coincides with the exact solution derived in \cite{BroAlm04}. 
We start from a quadratic Hamiltonian
\begin{equation}
H(\vct x) = \vct x\cdot \mat H \vct x
\label{quadham}
\end{equation}
with some symmetric matrix $\mat H$. The complex double phase space Hamiltonian,
\begin{equation}
\HH_c(\vct x,\vct y) = -2\vct x\cdot\mat H \mat J \vct y - \gamma\; \vct x\cdot \vct y
- \frac{i}{2} ~\Bigl[(\vct\lambda'\cdot\vct y)^2 + (\vct\lambda''\cdot\vct y)^2 \Bigr],
\end{equation}
induces the following double phase space dynamics:
\bea
\dot{\vct x}_\tau & = & \left(2 \mat J \mat H - \gamma\right) \vct x_\tau 
- i(\vct\lambda'\cdot\vct y_\tau)\; \vct\lambda'-i(\vct\lambda''\cdot\vct y_\tau)\; \vct\lambda'' \cr
\dot{\vct y}_\tau & = & \left(2 \mat H \mat J + \gamma\right) \vct y_\tau.
\eea
Notice that in this case, even though $\vct x_\tau$ is generally complex, 
this motion does not affect the trajectory $\vct y_\tau$, which remains real at all times.
Then the ``history'' of $(\vct y,t)$ is
\bea
\bar{\vct x}_\tau & = & - e^{-\gamma\tau}
\mat R_{\tau}\frac{\der S_0}{\der \vct y}
\left(e^{-\gamma t} \mat R_{t}^\top \vct y\right)
-i \int_0^\tau \;\left[(\vct\lambda'\cdot \vct y_u)\vct\lambda'+(\vct\lambda''\cdot \vct y_u)\vct\lambda''\right] ~du
\cr
\bar{\vct y}_\tau & = & e^{\gamma(\tau-t)}
\mat R_{t-\tau}^\top  \vct y,
\eea
in the notation of section \ref{complexd}.
Here $\mat R_{t}$ defines the classical propagation operator generated by the Hamiltonian (\ref{quadham}), that is
\begin{equation}
\mat R_{t} = \exp{\left(2\mat J \mat H t\right)},
\end{equation}
and $\mat R_t^\top$ stands for its transpose.
Because $\HH(\vct x,\vct y,t)$ is linear in $\vct x$, one has besides
\begin{equation}
-\bar{\vct x}_\tau \cdot \frac{\der {\bar{\vct y}_\tau}}{\der \tau} = \HH(\bar{\vct x}_\tau, \bar{\vct y}_\tau),
\end{equation}
so that, the expression of $S(\vct y,t)$ given by (\ref{S}) boils down to
\be
S(\vct y,t) 
=
S_0(e^{-\gamma t} \mat R_{t}^\top  \vct y) 
+ 
\frac{i}{2} \int_0^t 
 \left| \vct \lambda \cdot e^{\gamma(\tau-t)} \mat R_{t-\tau}^\top  \vct y \right|^2 ~d\tau .
\label{Squad}
\end{equation}
Hence, the full complex WKB solution can be written as
\begin{equation}
\chi (\vy,t) = \chi_0(e^{-\gamma t}\mat R_{t}^\top\;\vct y)\;\;  
\exp{\left( -\frac{1}{2\hbar} \int_0^t e^{2\gamma(t'-t)} \left| \vct \lambda 
\cdot \mat R_{t-t'}^\top\vct y\right|^2  ~dt' \right)}.
\label{exact-sol-chordy}
\end{equation}

Recalling our use of the same symbol, $\chi(\vct y)$, as a shorthand for $\chi(\vct \xi,t)=\chi(-\mat J \vct y,t)$,
together with (\ref{lamb}), we find that 
the complex WKB approximation coincides exactly with the exact solution in the quadratic case.
Indeed, the general solution derived in \cite{BroAlm04} was written in the following form 
\begin{equation}
\chi(\vct \xi,t) = \chi_0(e^{-\gamma t}\mat R_{-t}\;\vct \xi)\;\;  
\exp{\left( -\frac{1}{2\hbar} \int_0^t e^{2\gamma(t'-t)} \left| \vct l 
\cdot \mat R_{t'-t}\vct \xi\right|^2  ~dt' \right)},
\label{exact-sol-chord}
\end{equation}
where $\chi_0(\vct \xi)$ is any general initial chord function. 
To see that (\ref{exact-sol-chord}) is identical with (\ref{exact-sol-chordy}) one just need to notice that
\be
\mat R_{-t} \vct \xi = - \mat R_{-t} \mat J \vct y = 
- \mat J \left( \mat R_t^\top \vct y \right).
\end{equation}

Thus, the general picture is that of exact classical propagation of the chord function
(the same as the unitary evolution generated by the quadratic Hamiltonian),
multiplied by a Gaussian factor that progressively attenuates the contribution of long chords.

\section{Real Dynamics}
\label{realWKB}
\subsection{General analysis}

Complex WKB theory is more elegant than its real counterpart, in that the evolving complex action
accounts explicitly for the entire evolution of the wave function, 
whereas the real WKB action has to be supplemented by an evolving amplitude. 
Even so, the advantage of working with a more intuitive real phase space 
may predominate, whenever both theories lead to similar results.
It is important to note that, even though both sets of parameters $\vct \lambda'$ and $\vct \lambda''$
are real, the last integral in the general complex WKB solution (\ref{solutchigen}) 
will not be real in the case of a complex trajectory.
Thus it is not immediately clear that this still has the role of 
attenuating the amplitude of long chords as in the exact quadratic theory.
In contrast, the real WKB theory \cite{AlmBro07}, which is also exact in the quadratic case,
generalizes the quadratic exponent in (\ref{exact-sol-chord}) by a negative
{\it decoherence functional} with monotonical increasing modulus.
 
The necessity of the complex formalism in the description of unitary evolution arises 
when the amplitudes, $\alpha_j(\vct{\xi})=\alpha_j(-\mat J \vct y)$, in
\begin{equation}
\label{semichord}
\chi(\vct{\xi}) = \sum_j \; \alpha_j (\vct{\xi})\;e^{i S_j(\vct{\xi})/\hbar}, 
\end{equation} 
do not vary smoothly as compared to the complex oscillations.
In this situation, simple (real) stationary phase approximations are not allowed, 
i.e., when the variation of real part of the action is also divided by the small parameter, $\hbar$,
it is necessary to resort to full complex saddle point approximations. Where this difficulty is not
present, it was shown that both WKB theories lead to equivalent descriptions in the case of unitary evolution \cite{Hub03}.
In short, the evolution of the imaginary part of the complex action matches that of the logarithm of the
evolving real WKB amplitude.

The formal similarity between the Schr\"odinger equation and the Lindblad equation (\ref{dynchi_der_H}) 
for the chord function would permit us to immediately incorporate this equivalence, 
were it not for the imaginary part of the double Hamiltonian (\ref{compdouble}). 
Notice that the presence of the Lindblad dissipation coefficient, $\gamma$, 
in the real part of the double Hamiltonian (\ref{defHH}) is not a problem. 
The difficulty is that an initial real action cannot continue to be real, 
as it evolves classically due to a general complex Hamiltonian. 
The correct description of this feature is an advantage of the complex theory in the present context, which will now be examined.

A simple rederivation of the real space theory developed in \cite{AlmBro07} is to treat the imaginary part of the action
through classical perturbation theory. In the absence of the imaginary term, 
a real action defines a real Lagrangian manifold through (\ref{Lagman}), which evolves through real trajectories.
The first order effect of a perturbation of the Hamiltonian (whether real or complex) is to add a term,
\begin{equation}
\delta S_p = -i \int_0^t ~\delta\HH\left( \tilde{\vct x}_\tau,\tilde{\vct y}_\tau,\tau\right)~d\tau,
\end{equation}  
integrated over the (real) unperturbed trajectory, that we denote $(\tilde{\vct x}_\tau,\tilde{\vct y}_\tau)$ not to confuse it with the complex one $(\bar{\vct x}_\tau,\bar{\vct y}_\tau)$. In the present case,
\begin{equation} 
\delta\HH\left( {\vct x},{\vct y}\right)=
-\frac{1}{2} \left[(\vct \lambda' \cdot \vct y)^2+ (\vct \lambda'' \cdot \vct y)^2\right]
= -\frac{1}{2} \left|(\vct \lambda \cdot \vct y)\right|^2, 
\end{equation}
so that $\exp(i \delta S_p /\hbar)$
is just the decoherence term in the real WKB theory \cite{AlmBro07}: 
It damps out the amplitude of long chords, 
thus cancelling the fine oscillations of the Wigner function.

How crude is this approximation to the full complex theory? The way in which the classical
perturbation is rederived in Appendix \ref{app_realWKB} allows us to analyze this question.
There, we consider a general one-parameter family of Hamiltonians, so that henceforth
we replace $\delta\HH\left( {\vct x},{\vct y}\right)$ by $\alpha\;\delta\HH\left( {\vct x},{\vct y}\right)$
such that, in the present case, the parameter in Appendix \ref{app_realWKB} will be fixed as $\alpha =-i$. 
Applying this exact general treatment of a one parameter family of Hamiltonians,
the error in the perturbation theory, as compared to the exact action (\ref{exphase})
can thus be expressed as
\begin{eqnarray}
\Delta P= \delta S - \delta S_p = 
 \int_0^{-i} d\alpha' \int_0^{-t} d\tau\; 
[\delta\HH(\tilde{\vct x}_\tau(\alpha'), \tilde{\vct y}_\tau(\alpha'))
- \delta\HH(\tilde{\vct x}_\tau(0), \tilde{\vct y}_\tau(0))]
 \cr
= \int_0^{-i} d\alpha' \int_0^{-t} d\tau\; \frac{1}{2}
\left[\;(\vct \lambda' \cdot \tilde{\vct y}_\tau(\alpha'))^2 -(\vct \lambda' \cdot \tilde{\vct y}_\tau(0))^2
+\;(\vct \lambda'' \cdot \tilde{\vct y}_\tau(\alpha'))^2 -(\vct \lambda'' \cdot \tilde{\vct y}_\tau(0))^2\right].
\end{eqnarray} 
But according to (\ref{Hamilton}), $\dot{\vct y}$ is not directly affected
by the parameter $(\alpha=-i)$, so that this error grows slowly as $t$ is increased. 
Indeed, at any point in phase space the trajectory $\tilde{\vct y}_\tau$ has a velocity $\dot{\vct y}$
that is independent of $\alpha$, but the curvature of the trajectory is affected, because
\begin{equation}
\ddot{\vct y}(\alpha)= \ddot{\vct y}(0)+ 
\alpha \; (\vct\lambda' \cdot \vct y)\; \frac{\der^2 \HH}{\der \vct x^2}\; \vct\lambda'
+ \alpha \; (\vct\lambda'' \cdot \vct y)\; \frac{\der^2 \HH}{\der \vct x^2}\; \vct\lambda''.
\end{equation}
Thus, keeping constant the initial variable $\tilde{\vct y}_0(\alpha)=\vct y$ leads to the approximate trajectory,
\begin{equation}
\tilde{\vct y}_\tau(\alpha)\approx \vct y
+\frac{\alpha t^2}{2}\Bigl[(\vct\lambda' \cdot \vct y)\;\frac{\der^2 \HH}{\der \vct x^2}\;\vct\lambda'
+(\vct\lambda'' \cdot \vct y)\;\frac{\der^2 \HH}{\der \vct x^2}\;\vct\lambda''\Bigr],
\end{equation}
and the initial estimate of the error is
\begin{equation}
\Delta P\approx -\;\frac{\alpha^2 t^3}{6}\;\Bigl[(\vct\lambda' \cdot \vct y)^2\;
\left({\vct\lambda'}^T\frac{\der^2 \HH}{\der \vct x^2}\;\vct\lambda' \right)
+ (\vct\lambda'' \cdot \vct y)^2\;
\left({\vct\lambda''}^T\frac{\der^2 \HH}{\der \vct x^2}\;\vct\lambda'' \right)\Bigr].
\end{equation}

This first correction of the complex action is entirely real, because
the final parameter is $\alpha^2=i^2=-1$. If $\vct\lambda=\mat J \vct l$ is considered small, as is usually
assumed for Markovian evolution (the small coupling limit), 
then the error is $\GO(|\vct l|^4)$, while the decorerence functional is only
$\GO(|\vct l|^2)$. We also find that the decoherence functional grows linearly in time, 
while its error grows as $t^3$. A more subtle reason for the appropriateness of the real WKB approximation
is that the decoherence functional damps the contribution of all large chords, so that, for long times,
we are only interested in the classical region where $\vct y \approx 0$. 
 
\subsection{Real dynamics of the mixed propagator}
\label{dynmixedprop}

In \cite{AlmBro06} and \cite{AlmBro07} we introduced the mixed propagator $R_{\vct x}(\vct y,t)$, such that
\be
\chi(\vct y,t) = \int W_0(\vct x) R_{\vct x}(\vct y,t) ~d\vct x.
\label{mixedprop}
\ee
This propagator is a good candidate for applying the real dynamics because its initial expression is
\be
R_{\vct x}(\vct y,0) = \exp{\left(-\frac{i}{\hbar}\vct x\cdot \vct y\right)},
\label{defmixedprop}
\ee
which means that the initial action is purely real. Hence, the growth of the imaginary part of the action is initially perturbative, as will be verified in the following example. The important point is that this propagator can evolve any initial state, by using (\ref{mixedprop}).

The previous perturbative approach gives the following expression for this propagator,
\be
R_{\vct x}(\vct y,t) = \exp{\left(-\frac{i}{\hbar}\vct x\cdot \tilde{\vct y}_0 + \frac{i}{\hbar}\int_0^t\Bigl[\tilde{\vct x}_\tau \cdot \frac{\der \HH}{\der \vct x}(\tilde{\vct x}_\tau,\tilde{\vct y}_\tau,\tau) - \HH(\tilde{\vct x}_\tau,\tilde{\vct y}_\tau,\tau)\Bigr]~d\tau - \frac{1}{2\hbar}\int_0^t \Bigl[(\vct \lambda'\cdot \tilde{\vct y}_\tau)^2+ (\vct \lambda''\cdot \tilde{\vct y}_\tau)^2\Bigr]~d\tau \right)},
\label{perturbmixed}
\ee
where $(\tilde{\vct x}_\tau,\tilde{\vct y}_\tau)$ is the real history of $\vct y$ at time $t$ and with $\tilde{\vct x}_0 = \vct x$, governed by $\HH$ and not $\HH_c$. Hence $\tilde{\vct x}_\tau$ and $\tilde{\vct y}_\tau$ are real functions of $(\vct x, \vct y,t)$.

This expression has the advantage of separating the Lindbladian damping term from the oscillating phase, which is not the case for a general complex WKB. Therefore one can conclude that for short enough times, the chord function is fading exponentially fast for values of $\vct y$ which are outside a disk centred in $\vct 0$ and with radius $\sqrt{\hbar}$, as it does in the quadratic case \cite{BroAlm04}. Moreover, one can also conclude from (\ref{mixedprop}) that it is also true for any initial state, although with possible modulation of velocity since the damping term depends on the initial $\vct x$.

In the next section we treat the case of an initially Gaussian state, as an example where the above calculated propagator can be used to get a general formula. It relies on the fact that the contributing $\vct x$, in $R_{\vct x}(\vct y,t)$, are then localized around the Gaussian's centre.

\section{Dynamics of a Gaussian wave packet using the mixed propagator}

An initially localized wave packet will restrict the range of the required trajectories of (\ref{solutchigen}). For instance, if one starts with a Gaussian wave packet centred in $\vct X$, 
\be
W_0(\vct x) = \frac{1}{\pi\hbar}\exp{\left(-\frac{1}{\hbar}\left(\vct x - \vct X\right)^2 \right)},
\ee
then only trajectories whose starting point is within a distance $\sqrt{\hbar}$ from $\vct X$ will contribute to the integral. By expanding the generic trajectory around the central one $(\bar{\vct x}^X_\tau,\bar{\vct y}^X_\tau)=\left(\bar{\vct x}_\tau(\vct X,\vct y,t), \bar{\vct y}_\tau(\vct X,\vct y,t)\right)$, corresponding to the evolution of the the maximum $\vct X$ of the initial Gaussian, one can perfom the explicit integration of (\ref{mixedprop}).

If we call $S(\vct X, \vct y,t) = S'(\vct X, \vct y,t)+ i S''(\vct X, \vct y,t)$ the action such that $R_{\vct X}(\vct y,t)=\exp{\frac{i}{\hbar}S(\vct X, \vct y,t)}$, then $R_{\vct X+\vct x'}(\vct y,t)$ can be expanded as
\be
R_{\vct X+\vct x'}(\vct y,t)=\exp{\left(\frac{i}{\hbar}S(\vct X, \vct y,t)+\frac{i}{\hbar}\frac{\der S}{\der \vct x}(\vct X, \vct y,t)\cdot \vct x' + \frac{i}{2\hbar}\vct x'\cdot \frac{\der^2 S}{\der \vct x^2}(\vct X,\vct y,t) \vct x' + \frac{i}{\hbar}\GO(\vct x'^3) \right)}.
\label{expandperturbmixed}
\ee
By interpreting $S$ as the generating function of the complex manifold driven by the complex Hamiltonian $\HH_c$, we show in appendix \ref{app_derS} that
\be
\frac{\der S}{\der \vct x}(\vct X, \vct y,t) = -\bar{\vct y}_0,
\ee
where $\bar{\vct y}_0$ is the initial $\vct y$ coordinate of the complex trajectory $(\vct X,\bar{\vct y}_0,0)\rightarrow(\vct x_{\mathcal M}(\vct y,t),\vct y,t)$, constrained to start at $\vct x = \vct X$ and to stop at $\vct y$ in a time $t$.

Then (\ref{mixedprop}) gives
\be
\chi(\vct y,t) \simeq \frac{1}{\pi\hbar} \int \exp{\left(-\frac{1}{\hbar}\vct x'^2 + \frac{i}{\hbar}S(\vct X, \vct y,t)-\frac{i}{\hbar}\bar{\vct y}_0\cdot \vct x' + \frac{i}{2\hbar}\vct x'\cdot \frac{\der \bar{\vct y}_0}{\der \vct x} \vct x' \right)} ~d\vct x',
\ee
and then
\be
\chi(\vct y,t) \simeq \frac{1}{\sqrt{\det{\left(1-\frac{i}{2}\frac{\der \bar{\vct y}_0}{\der \vct x}\right)}}} \exp{\left(\frac{i}{\hbar}S(\vct X, \vct y,t) - \frac{1}{\hbar}\bar{\vct y}_0 \cdot \left( 1 - \frac{i}{2} \frac{\der \bar{\vct y}_0}{\der \vct x} \right)^{-1} \bar{\vct y}_0 \right)},
\label{quadchi}
\ee
where $\bar{\vct y}_0$ is a complex valued function of $(\vct X, \vct y,t)$, hence a function of $(\vct y,t)$ once the initial state is fixed.

The analysis holds as long as $S(\vct x,\vct y,t)$ can be approximated by its quadratic expansion in $\vct x$, that is, when the dynamics is sufficiently regular, or time sufficiently small. For a chaotic dynamics one expects exponentially diverging trajectories which may make expansion (\ref{expandperturbmixed}) irrelevant for large times.

If one uses the real WKB theory instead of the complex WKB theory, then the imaginary part of $\bar{\vct y}_0$ can be interpreted as a perturbation of the origin $\tilde{\vct y}_0$ of the real trajectory $(\tilde{\vct x}^X_\tau,\tilde{\vct y}^X_\tau)$:
\be
\frac{\der S}{\der \vct x}(\vct X, \vct y,t) \simeq -\tilde{\vct y}_0 + 
i\int_0^t \left(\vct \lambda \cdot \tilde{\vct y}_\tau\right)\frac{\der \tilde{\vct y}_\tau}{\der \vct x}\vct \lambda ~d\tau.
\ee
In the following example, both theories give the same result.

\section{A cubic example}

Let us consider the special case of the Hamiltonian $\op H(p,q)=\op p^3$ and a linear Lindblad operator, $\op L = l \op p$. 
The exact solution can be obtained, which allows us to check on our different levels of approximation.
The Lindblad equation can be written
\begin{equation}
\frac{\der \op \rho}{\der t} = -\frac{i}{\hbar}\Bigl[ \op H, \op \rho \Bigr]
+ \frac{l^2}{2\hbar}\left(
2\op p \op \rho \op p - \op p^2 \op \rho - \op \rho \op p^2.
\right)
\end{equation}
In the $p$ representation we then have
\begin{equation}
\frac{\der }{\der t}\langle p' |\op \rho (t)| p'' \rangle  
= -\frac{i}{\hbar}({p'}^3 - {p''}^3) \langle p' |\op \rho(t) | p'' \rangle
- \frac{l^2}{2\hbar} (p' - p'')^2 \langle p' |\op \rho(t) | p'' \rangle,
\end{equation}
which gives the solution
\begin{equation}
\langle p' |\op \rho(t) | p'' \rangle = \langle p' |\op \rho(0) | p'' \rangle 
\exp{\left(-\frac{i}{\hbar}({p'}^3 - {p''}^3)t
- \frac{l^2}{2\hbar} (p' - p'')^2 t
\right)}.
\end{equation}
The general expression for the chord function is then
\begin{eqnarray}
\chi(\vct y,t) = \int dp~ \langle p + \frac{y_q}{2} | \op \rho(t) | p - \frac{y_q}{2} \rangle 
\exp{\left( -\frac{i}{\hbar}p y_p\right)} \cr 
=\exp\left(-\frac{it}{4\hbar}y_q^3-t\frac{l^2 }{2\hbar}y_q^2\right) 
\int dp~ \langle p + \frac{y_q}{2} | \op \rho(0) | p - \frac{y_q}{2} \rangle 
\exp{\left( -\frac{i}{\hbar}(y_p \;p+t\; y_q \;3p^2 )\right)},
\label{intchord}
\end{eqnarray}
i.e. the original chord function is convoluted with an evolving Gaussian 
and then multiplied by another time dependent factor.

The complex WKB approximation for this evolution is 
based on the classical dynamics (\ref{Hamilton})
of the complex double phase space Hamiltonian (\ref{defHH}), which is here equal to
\begin{equation}
\HH_c(p,q,y_p,y_q)=3p^2y_q + \frac{y_q^3}{4} - \frac{i}{2}(ly_q )^2.
\end{equation}
It generates the following trajectories,
\begin{eqnarray}
(\bar y_q)_t & = & (y_q)_0  = y_q \cr
(\bar y_p)_t & = & (y_p)_0 - 6 p_0 y_q t \cr
\bar q_t & = & q_0 + (3p_0^2 + \frac{3}{4}y_q^2)t + i l^2 y_q t\cr
\bar p_t & = & p_0.
\label{traj}
\end{eqnarray}
One should note that in this simple example the chord motion, $\bar{\vct y}_t$,
is independent of the trajectory of the centre variable, 
$\bar{\vct x}_t= (\bar q_t, \bar p_t)$, just as occurs for quadratic systems.
Hence, the evolved chord will remain real if it is real to start with.

Given the initial action, $S_0(\vct y)$, its evolution is obtained explicitly from (\ref{S})
as 
\begin{equation}
S(\vct y,t)  = S_0(y_p + 6 p_0 y_q t,y_q)
+ t \left( 3p_0^2 y_q - \frac{y_q^3}{4} + \frac{i}{2} (l y_q)^2\right),
\end{equation}
leading to the complex WKB approximation: 
\begin{eqnarray}
\chi_{WKB}(\vct y,t) = \exp{ \frac{i}{\hbar}\left(S_0({\vct y}_0)
+ t\; y_q \;3p_0^2- \frac{ty_q^3}{4} + \frac{it}{2} (l y_q)^2 )\right)},
\label{WKBEX}
\end{eqnarray}

This can now be compared with the approximate saddle point evaluation of (\ref{intchord}).
Note that the initial density operator in the $p$-representation, corresponding to (\ref{simple}) with the initial action $S_0(y)$, will also have a standard semiclassical form. 
It is just the Fourier inverse of (\ref{intchord}) for $t=0$, 
with an action $\sigma_0(p', p'')$ (a symmetrized Legendre transform of $S_0(\vct y)$),
such that 
\begin{equation}
\frac{\der\sigma_0}{\der p'}= -q'_0\;\;,\;\;\frac{\der\sigma_0}{\der p''}= q''_0.
\end{equation}
Then the integral expression (\ref{intchord}) for the evolving chord function 
can be performed by the saddle point method.
Thus the stationary point, $p_0$, of the exponent of (\ref{intchord}) 
is selected by the equation
\begin{equation}
q_0(p_0+y_q/2)-q_0(p_0-y_q/2)= y_p + 6t y_q p_0 = (y_p)_0,
\label{stat}
\end{equation}
where, in the last equality, we have followed the $y_p$-component 
of the classical trajectory in (\ref{traj}) backwards in time.
This trajectory depends on $p_0$, a constant, according to (\ref{traj}). 
Hence, the full approximation for the evolving chord function becomes
\begin{eqnarray}
\chi_{SP}(\vct y,t) = \sqrt{\frac{\pi\hbar}{\frac{\der^2 \sigma_0}{\der p^2}(p+\frac{y_q}{2},p-\frac{y_q}{2})+ 3i y_q t }}
\exp{\left(- \frac{ity_q^3}{4\hbar}-\frac{t\;l^2y_q^2 }{2\hbar}+\frac{i}{\hbar}\sigma_0(p+\frac{y_q}{2},p-\frac{y_q}{2})+\frac{i}{\hbar} t\; y_q \;3p_0^2 \right)},
\label{sp_intchord}
\end{eqnarray}
Thus the phase of $\chi_{SP}(\vct y,t)$ coincides with that of $\chi_{WKB}(\vct y,t)$.

In some cases this phase is exact. 
Consider, the simple example of a ``Schr\"odinger cat state'', composed of the superposition of a pair of coherent states, with the wave function
\begin{eqnarray}
\psi(q) = \frac{1}{(\pi\hbar)^{1/4}}
\exp{\left(-\frac{(q-Q)^2}{2\hbar}+\frac{iP q}{\hbar}
\right)} ,
\end{eqnarray}
and different parameters, $\vct X_a=(P_a, Q_a)$ and $\vct X_b=(P_b, Q_b)$.
The density operator, corresponding to $|\psi_a\rangle + |\psi_b\rangle$, 
has two diagonal terms and two off-diagonal terms:
\begin{equation}
\op \rho = \frac{1}{2}|\psi_a\rangle \langle\psi_a| + 
\frac{1}{2}|\psi_b\rangle \langle\psi_b| + 
\frac{1}{2}|\psi_a\rangle \langle\psi_b| 
+ \frac{1}{2}|\psi_b\rangle \langle\psi_a|.
\label{cat}
\end{equation}
We are interested here in the sum of the pair of non-local terms, 
$\op \rho_{ab}=|\psi_a\rangle \langle\psi_b|$ and its complex conjugate. 
One can notice that the two first ``classical'' coherent terms can be retrieved by setting $a=b$.

Inserting the $p$-representation of the initial density matrix,
\begin{equation}
\langle p' |\op \rho_{ab}(0) | p'' \rangle = 
\frac{1}{\sqrt{\pi\hbar}}\exp{\left( \frac{i}{\hbar}(p'Q_a-p''Q_b) -\frac{1}{2\hbar}(p'-P_a)^2
-\frac{1}{2\hbar}(p''-P_b)^2\right)},
\end{equation}
into (\ref{intchord}), the integral can be performed to give the exact chord function evolution:
\begin{eqnarray}
\chi_{ab}(\vct y,t) & = &\frac{1}{\sqrt{1+3i y_q t}}
\exp{
\left( \frac{1}{4\hbar}
\frac{4P^2 - (y_p+\Delta Q)^2 - 4iP(y_p+\Delta Q)}{1+3ity_q} \right. } \cr
~ & ~ & { \left.  
-\frac{1}{4\hbar}(y_q-\Delta P)^2 - \frac{1}{2\hbar}l^2 t y_q^2 - \frac{1}{\hbar}P^2 - \frac{i}{\hbar}Qy_q - \frac{i}{4\hbar}ty_q^3
\right)},
\label{exactchordcub}
\end{eqnarray}
where we have defined:
\bea
P & = & \frac{P_a+P_b}{2} \cr
Q & = & \frac{Q_a+Q_b}{2} \cr
\Delta P & = & P_b - P_a \cr
\Delta Q & = & Q_b - Q_a.
\label{param}
\eea

The initial action obtained from setting $t=0$ is
\begin{equation}
S_0(\vct y) =  - Qy_q - P(y_p + \Delta Q) +
\frac{i}{4}(y_p + \Delta Q)^2 + \frac{i}{4}(y_q - \Delta P)^2, 
\end{equation}
so that
\begin{eqnarray}
p_0 & = & - \frac{\der S_0}{\der y_p}(\vct y_0) = P + 
\frac{i}{2}(y_p + 6 p_0 y_q t + \Delta Q)
\cr
q_0 & = & - \frac{\der S_0}{\der y_q}(\vct y_0) = Q +
\frac{i}{2}(y_q - \Delta P).
\end{eqnarray}
Then equation (\ref{S}) gives
\bea
S(\vct y,t)  = 
S_0(y_p + 6 p_0 y_q t,y_q)
+ t \left( 3p_0^2 y_q - \frac{y_q^3}{4} - \frac{i}{2} (l y_q)^2\right) \cr
 =  
-Qy_q - P(y_p+6p_0ty_q) + \frac{i}{4}(y_p+6p_0ty_q+\Delta Q)^2 
+ t \left( 3p_0^2 y_q - \frac{y_q^3}{4} - \frac{i}{2}(ly_q)^2\right). 
\label{approxS}
\eea
Here,
\be
p_0 = \frac{P + \frac{i}{2}(y_p + \Delta Q)}{1+3iy_qt},
\end{equation}
corresponds to the explicit evaluation of the stationary point defined by (\ref{stat}).
Therefore, both $\chi_{SP}(\vct y,t)$ and $\chi_{WKB}(\vct y,t)$ have the exact phase,
being that the former also has the exact amplitude.

This example of an initially localized wavepacket is a case where the real WKB theory becomes a crude approximation, because the initial action is then complex and, hence, the contributing trajectory is also complex.

However the simple classical nonquadratic evolution in this section does not lead to a complex trajectory in (\ref{traj}), if it is initially real, i.e., for an initial real action. In this case, the saddle point evaluation of (\ref{intchord}) becomes equivalent to the stationary phase method, leading
to the real WKB result. In other words, the first order perturbation of the phase, which was studdied in section \ref{realWKB}, gives the full correction with respect to the complex WKB phase.

In particular, the real WKB theory can be used to propagate the real initial action of the mixed propagator $R_{\vct x}(\vct y,t)$ presented in section \ref{dynmixedprop}, with its chord space representation (\ref{defmixedprop}). It corresponds to a density operator given by
\be
\langle p'+\frac{y_q}{2} | \op \rho | p'-\frac{y_q}{2} \rangle = \delta(p'-p)\exp{\left(-\frac{i}{\hbar}q\;y_q\right)},
\ee
with $\vct x = (p,q)$. Then, the integral in (\ref{intchord}) can again be performed exactly to obtain:
\begin{equation}
{R}_{\vct x}(\vct y, t)=\exp\left(-\frac{it}{4\hbar}y_q^3-t\frac{l^2 }{2\hbar}y_q^2\right) \exp{\left( -\frac{i}{\hbar}y_q\;q - \frac{i}{\hbar}y_p \;p-\frac{i}{\hbar}t\; y_q \;3p^2 \right)}.
\label{Rx_realWKB}
\end{equation}
This expression coincides exactly with (\ref{perturbmixed}), which can be obtained explicitely from (\ref{traj}) by setting $\vct l = \vct 0$. Thus, in this case both the real and the complex WKB evolutions are exact.

This is all the more interesting, in that $R_{\vct x}(\vct y,t)$ can then be used to propagated any initial state. For instance, with the ``Schr\"odinger cat state'', one gets
\be
\chi_{ab}(\vct y,t) = \int W_0(\vct x){R}_{\vct x}(\vct y, t) ~d\vct x,
\label{Rxpropgauss}
\ee
with
\be
W_0(\vct x) = \exp{\left(\frac{p^2+q^2}{\hbar}+\frac{i}{\hbar}2y_p\;p\right)},
\ee
which again gives (\ref{exactchordcub}). This shows that the real WKB method, although inacurate for a general initial state, can still be used indirectly through $R_{\vct x}(\vct y,t)$.

Notice that expression (\ref{Rxpropgauss}) then also coincides with expression (\ref{quadchi}), which is not so surprising as the action of ${R}_{\vct x}(\vct y, t)$, in (\ref{Rx_realWKB}), is actually quadratic in $\vct x$.

\section{Conclusion}

We have shown that to leading order in $\hbar$, the chord function, or characteristic function, of a generic state will evolve according to a simple expression (\ref{solutchigen}) which is mainly determined by a generalized Hamiltonian dynamics taking place in a complex double phase space $(\vct x,\vct y)$. The subspace $\vct y=\vct 0$ corresponds to the ordinary classical phase space dynamics, whereas nonzero $\vct y$ accounts for quantum (nonclassical) correlations. 

When the classical trajectories used to build the phase of this expression are real, then the Lindbladian term is growing negatively and contributes to an exponential damping of the values of $\vct y$ which are beyond the disk of radius $\sqrt{\hbar}$ around $\vct 0$. This can be recognized as the region containing all the classical information about the state. The overall features of this evolution are certainly present in the case of a quadratic Hamiltonian and linear Lindblad operators, as derived for instance in \cite{BroAlm04}, and fulfills our handwaving intuition that ``decoherence drives the system back into classical dynamics''. 

We have here shown that this scenario also holds for general Hamiltonians, when the action of the initial chord function is real, and time short enough, so that the imaginary part of classical dynamics grows. Furthermore, it can be extended to any initial state by using the mixed propagator $R_{\vct x}(\vct y,t)$, which fulfills the requirement for initially real action. 

For short times, the evolution of the initially localized components of a Shr\"odinger cat state, follow a simplified dynamics that may be understood without the full recourse to complex orbits. This is as true as of unitary evolution and the same methods employed in \cite{Heller,Littlej} can be generalized to a real double phase space. This approach has been presented in \cite{BroAlm08}

For larger times the real and imaginary parts of the phase of the chord function become intertwined, and the evolution becomes difficult to unravel. However the decoherence time is known to be very small, and if complete decoherence is reached before the limit of perturbative validity, then the Wigner function becomes positive, and hence it will stay positive, because the Lindblad equation preserves this property.

\appendix

\section{Asymptotic expansion of the Hamiltonian operator}
\label{app_devH}

One can verify easily that
\begin{eqnarray}
\left( \frac{1}{\lambda} \frac{\der}{\der p} \right)^m
\left( \frac{1}{\lambda} \frac{\der}{\der q} \right)^n \exp{\lambda S} & = &
\Biggl[
\left( \frac{\der S}{\der p} \right)^m \left( \frac{\der S}{\der q} \right)^n
+ \frac{1}{\lambda} 
\left( 
\frac{m(m-1)}{2} \frac{\der^2 S}{\der p^2} \left( \frac{\der S}{\der p} \right)^{m-2}
\left( \frac{\der S}{\der q} \right)^n  
\right. 
 \cr 
~ & ~ & 
\left.
+ mn \frac{\der^2 S}{\der p \der q} \left( \frac{\der S}{\der p} \right)^{m-1} \left( \frac{\der S}{\der q} \right)^{n-1} + \frac{n(n-1)}{2} \frac{\der^2 S}{\der q^2} \left( \frac{\der S}{\der p} \right)^{m}
\left( \frac{\der S}{\der q} \right)^{n-2}
\right) 
 \cr
~ & ~ & 
+ \GO(\frac{1}{\lambda^2})
\Biggr] \exp{\lambda S}.
\end{eqnarray}
From there one can generalize and write
\begin{equation}
\HH\left( -\frac{\hbar}{i}\frac{\der}{\der \vct y}, \vct y \right)
\exp{\left(\frac{i}{\hbar}S \right)} = \Biggl[
\HH\left(-\frac{\der S}{\der \vct y}, \vct y \right) + \frac{\hbar}{2i}\mathop{Tr}{\{\frac{\der^2 \HH}{\der \vct x^2}\left(-\frac{\der S}{\der \vct y}, \vct y \right)
\frac{\der^2 S}{\der \vct y^2}\}} + \GO(\hbar^2)
\Biggr]\exp{\left(\frac{i}{\hbar}S \right)}
\end{equation}

\section{Proof of the closeness of the differential form $\delta a$}
\label{app_closedform}

We prove that the form
\begin{equation}
\delta a(\vct y,t)=-\vct x_{\mathcal M}(\vct y,t)
\cdot d\vct y - \HH(\vct x_{\mathcal M}(\vct y,t),\vct y,t)dt
\end{equation}
is a closed form by showing that it obeys Schwartz equalities
\begin{equation}
\frac{\der }{\der t}\vct x_{\mathcal M}(\vct y,t) = \frac{\der}{\der \vct y} \HH(\vct x_{\mathcal M}(\vct y,t),\vct y,t),
\label{Schwartz}
\end{equation}
and
\begin{equation}
\frac{\der p_{\mathcal M}(\vct y,t)}{\der y_q} = \frac{\der q_{\mathcal M}(\vct y,t)}{\der y_p}.
\label{Schwartz2}
\end{equation}

The point $(\vct x_{\mathcal M}(\vct y,t),\vct y)$ is on a particular trajectory $(\bar{\vct x}_\tau,\bar{\vct y}_\tau)$ of the manifold ${\mathcal M}$ with
\begin{eqnarray}
\bar{\vct x}_0 & = & \vct x_{\mathcal M}(\vct y,t) \cr
\bar{\vct y}_0 & = & \vct y \cr
\end{eqnarray}
and
\begin{eqnarray}
\dot{\bar{\vct x}}_\tau & = & 
\frac{\der \HH}{\der \vy}\left( \bar{\vct x}_\tau,\bar{\vct y}_\tau,t+\tau \right) \cr
\dot{\bar{\vct y}}_\tau & = &
- \frac{\der \HH}{\der \vx}\left( \bar{\vct x}_\tau,\bar{\vct y}_\tau,t+\tau \right).
\label{dery}
\end{eqnarray}

One also has, of course,
\begin{equation}
\vct x_{\mathcal M}(\bar{\vct y}_\tau,t+\tau) = \bar{\vct x}_\tau.
\label{xMxtau}
\end{equation}
Now we can write the following sequence of equalities
\begin{eqnarray}
\vct x_{\mathcal M}(\vct y,t+d\tau) - \vct x_{\mathcal M}(\vct y,t) 
& = &
\vct x_{\mathcal M}(\vct y,t+d\tau) - \vct x_{\mathcal M}(\vct y + \dot{\bar{\vct y}}_0d\tau,t+d\tau)
+ \vct x_{\mathcal M}(\vct y + \dot{\bar{\vct y}}_0d\tau,t+d\tau) - \vct x_{\mathcal M}(\vct y,t) \cr
~ & = &
- \frac{\der \vct x_{\mathcal M}}{\der \vct y}(\vct y,t+d\tau)\dot{\bar{\vct y}}_0d\tau
+ \bar{\vct x}_\tau  - \bar{\vct x}_0  + \GO(d\tau^2) \cr
~ & = & 
\frac{\der \vct x_{\mathcal M}}{\der \vct y}(\vct y,t+d\tau)\frac{\der \HH}{\der \vx} ( \bar{\vct x}_0,\bar{\vct y}_0,t)d\tau + \frac{\der \HH}{\der \vct y}( \bar{\vct x}_0,\bar{\vct y}_0,t ) d\tau + \GO(d\tau^2) \cr
~ & = & 
\frac{\der \vct x_{\mathcal M}}{\der \vct y}(\vct y,t+d\tau)\frac{\der \HH}{\der \vx} ( \vct x_{\mathcal M}(\vct y,t),\vct y,t)d\tau + \frac{\der \HH}{\der \vct y}( \vct x_{\mathcal M}(\vct y,t),\vct y,t ) d\tau + \GO(d\tau^2) \cr
~ & = & 
\frac{\der }{\der \vct y} \HH (\vct x_{\mathcal M}(\vct y,t),\vct y,t)d\tau + \GO(d\tau^2).
\end{eqnarray}
This last equality shows (\ref{Schwartz}).

On the other hand, at $t=0$ we have
\begin{equation}
\frac{\der p_{\mathcal M}}{\der y_q}(\vct y,0) = -\frac{\der}{\der y_q}
\frac{\der A_0}{\der y_p}(\vct y) = -\frac{\der}{\der y_p}
\frac{\der A_0}{\der y_q}(\vct y) = \frac{\der q_{\mathcal M}}{\der y_p}(\vct y,0),
\end{equation}
which sets (\ref{Schwartz2}) at $t=0$. Now one just has to notice that $\displaystyle \frac{\der p_{\mathcal M}}{\der y_q}(\bar{\vct y}_\tau,t+\tau) - \frac{\der q_{\mathcal M}}{\der y_p}(\bar{\vct y}_\tau,t+\tau)$ is time independent:
\begin{eqnarray}
\frac{\der}{\der \tau}
\left(  \frac{\der p_{\mathcal M}}{\der y_q}(\bar{\vct y}_\tau,t+\tau) 
- \frac{\der q_{\mathcal M}}{\der y_p}(\bar{\vct y}_\tau,t+\tau)  \right)
& = &
\frac{\der}{\der y_q} \frac{\der }{\der \tau} \bar p_\tau 
- \frac{\der}{\der y_p} \frac{\der }{\der \tau} \bar q_\tau   \cr
~ & = &
\frac{\der}{\der y_p} 
\frac{\der \HH }{\der y_q}(\bar{\vct x}_\tau,\bar{\vct y}_\tau,t+\tau) 
- \frac{\der}{\der y_q} 
\frac{\der \HH }{\der y_p}(\bar{\vct x}_\tau,\bar{\vct y}_\tau,t+\tau)
\cr
~ & = & 0,
\end{eqnarray}
which shows the statement. We used (\ref{dery}) and (\ref{xMxtau}).

Therefore, since $\mathcal M$ is simply connected, $\delta a(\vct y,\tau)$ is the exact differential of a function $a(\vct y,\tau)$ such that 
\begin{eqnarray}
\frac{\der a}{\der \vct y}(\vct y,\tau) 
& = & - \vct x_{\mathcal M}(\vct y,\tau) \cr
\frac{\der a}{\der \tau}(\vct y,\tau) & = & 
- \HH(\vct x_{\mathcal M}(\vct y,\tau),\vct y,\tau).
\label{generat}
\end{eqnarray}
Hence one has
\begin{equation}
\frac{\der a}{\der \tau}(\vy,\tau) = - \HH\left( - \frac{\der a}{\der \vy}(\vct y,\tau),\vct y ,\tau\right),
\label{asolut}
\end{equation}
so that $A(\vct y,t) = a(\vct y, t)$ is the required solution.

\section{Real WKB theory as a perturbation of the complex one}
\label{app_realWKB}

Let us consider a familly of Hamiltonians, $\HH(\vct x,\vct y; \alpha)$, where the parameter $\alpha$
may be real or complex. (It can also have more than one dimension.) It is then possible to reinterpret
$\HH$ as a single Hamiltonian in an expanded phase space: $(\vct x,\vct y; \alpha, \beta)$ in which
$\beta$ is a hidden variable. It follows that $\alpha$ is a constant of the motion and the projection
of each trajectory onto the $(\vct x,\vct y)$ hyperplane coincides exactly with the trajectory 
of the original family of Hamiltonians for its specific value of $\alpha$.

Consider now the root search problem for finding, given any value of $\alpha$, 
a trajectory that starts at the Lagrangian surface $\vct x= \partial S_0/\partial\vct y$ 
and arrives at specified  value of $\vct y$. Varying the parameter continuously in the
interval $[0,\alpha]$, we obtain a one parameter family of trajectory segments, 
which form a finite 2-D strip in the enlarged phase space. This can be fully specified
by choosing $\beta=0$ as the arrival point for the extra coordinate (for all $\alpha$) 
and its initial point to be
\begin{equation}
\beta_0(\alpha) = \int_0^{-t} d\tau\;\; \dot \beta_\tau = 
\int_0^{-t} d\tau\;\; \frac{\partial \HH}{\partial \alpha}(\tilde{\vct x}_\tau(\alpha), \tilde{\vct y}_\tau(\alpha); \alpha).
\label{beta0}
\end{equation}  

According to the Poincare-Cartan theorem \cite{Arnold:book},
\begin{equation}
\oint [\vct x \cdot d\vct y + \beta \cdot d\alpha - \HH \; d\tau] = 0
\label{PCthm}
\end{equation} 
for any reducible circuit on a 2-D surface that is spanned by trajectories. 
So we can build this circuit from four segments: i. The trajectory that
travels from the original Lagrangian surface and $\beta_0(\alpha)$ to $(\vct y= const, \beta=0)$ 
in the time $t$; ii. a path $\alpha\rightarrow 0$, 
with all other extended phase space variables and the time held fixed;
iii. the trajectory that reverses trajectory (i.), but with $\alpha=0$;
iv. The circuit is closed (in zero time) by path along the original Lagrangian surface,
while increasing $\alpha$ back to its original value, such that $\beta=\beta_0(\alpha)$,
as defined by (\ref{beta0}). Then (\ref{PCthm}) is rewritten as
\begin{eqnarray}
S_0(\vct y(\alpha)) + 
\int_0^t [\tilde{\vct x}(\alpha) \cdot \dot{\tilde{\vct y}}(\alpha) 
- \HH(\tilde{\vct x}_\tau(\alpha), \tilde{\vct y}_\tau(\alpha))]\; d\tau \cr 
=S_0(\vct y(0)) + \int_0^t 
[\tilde{\vct x}(0) \cdot \dot{\tilde{\vct y}}(0) - \HH(\tilde{\vct x}_\tau(0), \tilde{\vct y}_\tau(0))]\; d\tau
-\int_0^{\alpha}\; \beta_0(\alpha)\; d\alpha.  
\end{eqnarray}
Hence, the difference in the final action is exactly
\begin{equation}
\delta S(\alpha)= S(\vct y, t; \alpha)- S(\vct y, t;0) = 
 \int_0^{\alpha} d\alpha' \int_0^{-t} d\tau\; 
\frac{\partial \HH}{\partial \alpha'}(\tilde{\vct x}_\tau(\alpha'), \tilde{\vct y}_\tau(\alpha'); \alpha'). 
\label{exphase}
\end{equation}

For small enough $\alpha$, this may be approximated as
\begin{equation}
\Delta S(\alpha)=  \int_0^{-t} d\tau\; 
[\HH(\tilde{\vct x}_\tau(0),\tilde{\vct y}_\tau(0); \alpha)-\HH(\tilde{\vct x}_\tau(0),\tilde{\vct y}_\tau(0); 0)]
=\alpha\int_0^{-t} d\tau\;\delta\HH(\tilde{\vct x}_\tau(0),\tilde{\vct y}_\tau(0)) , 
\end{equation}
if $\HH(\vct x,\vct y; \alpha)=\HH(\vct x,\vct y)+ \alpha\;\delta\HH(\vct x,\vct y)$.
This is just the first order classical perturbation: One integrates the change in the Hamiltonian
along the unperturbed trajectory.

\section{Derivative of the action of the mixed propagator with respect to the centre space variable}
\label{app_derS}

For every $\vct x_0$ we define a manifold $\mathcal M(\vct x_0)$ by propagating the manifold $\vct x = \vct x_0$ in the double phase space driven by the complex Hamiltonian $\HH_c$. The coordinates of the manifold are $(\vct x_{\mathcal M}(\vct x_0, \vct y,t),\vct y,t)$, thus parametrized by $(\vct y,t)$.
We want to evaluate the infinitesimal difference $S(\vct x_0+d\vct x_0,\vct y,t) -S(\vct x_0,\vct y,t)$, where the variable $\vct x_0$ corresponds to the initial manifold, the variable $\vct y$ to the final observation point, and $t$ to the time of evolution. We have
\be
S(\vct x_0,\vct y,t) = - \vct x_0\cdot \bar{\vct y}_0 - \int_0^t \left(\bar{\vct x}_\tau\cdot \dot{\bar{\vct y}}_\tau - \HH_c( \bar{\vct x}_\tau,\bar{\vct y}_\tau,\tau)\right)~d\tau , 
\ee
where $(\bar{\vct x}_\tau,\bar{\vct y}_\tau)=(\vct x_{\mathcal M}(\vct x_0,\bar{\vct y}_\tau,\tau),\bar{\vct y}_\tau)$ is the Hamiltonian trajectory which starts at $(\vct x_0,\bar{\vct y}_0)$ at time $\tau = 0$ and ends at $\vct y$ at time $\tau =t$. On the other hand, one has
\be
S(\vct x_0+d\vct x_0,\vct y,t) = - (\vct x_0 + d\vct x_0)\cdot \bar{\vct y}_0  - \int_{\bar{\vct y}_0,0}^{\vct y,t} \vct x_{\mathcal M}(\vct x_0,\vct y,\tau) \cdot d\vct y + \HH_c( x_{\mathcal M}(\vct x_0,\vct y,\tau),\vct y,\tau)~d\tau ,
\ee
where we used the lagrangian nature of $\mathcal M(\vct x_0+d\vct x_0)$ which gives us the freedom to start from the point $(\vct x_0 + d\vct x_0,\bar{\vct y}_0)$ at $\tau=0$, which is indeed a point of the manifold, and follow any path which arrives at $\vct y$ on time $t$. In particular, one can chose the path $(\vct x_{\mathcal M}(\vct x_0+d\vct x_0,\bar{\vct y}_\tau,\tau),\bar{\vct y}_\tau,\tau)$, which follows ``vertically'' the $\mathcal M(\vct x_0)$ trajectory, but which is not a Hamiltonian trajectory on $\mathcal M(\vct x_0 + d\vct x_0)$. Then one can see that
\begin{eqnarray}
S(\vct x_0+d\vct x_0,\vct y,t) & = & S(\vct x_0,\vct y,t) - d\vct x_0\cdot \bar{\vct y}_0  \cr
~ & ~ & - \int_0^t \left(\frac{\der \vct x_{\mathcal M}}{\der \vct x_0}(\vct x_0,\bar{\vct y}_\tau,\tau)~d\vct x_0 \cdot\dot{\bar{\vct y}}_\tau  + \frac{\der \vct x_{\mathcal M}}{\der \vct x_0}(\vct x_0,\bar{\vct y}_\tau,\tau)~d\vct x_0\cdot\frac{\der \HH_c}{\der \vct x}(\bar{\vct x}_\tau,\bar{\vct y}_\tau ,\tau)\right) ~d\tau \cr
~ & = & S(\vct x_0,\vct y,t) - d\vct x_0\cdot \bar{\vct y}_0,
\end{eqnarray}
so that
\be
\frac{\der S}{\der \vct x}(\vct x_0,\vct y,t) = -\bar{\vct y}_0.
\ee

%\ack
\begin{acknowledgments}

Partial financial support from Millenium Institute of Quantum
Information, FAPERJ, PROSUL, CNPq and CAPES-COFECUB is gratefully acknowledged.

\end{acknowledgments}

\end{document}